\documentclass[aps,pra,10pt,tightenlines,longbibliography,notitlepage,amsmath,twocolumn,floatfix,superscriptaddress,eqsecnum]{revtex4-1}
\usepackage[dvipsnames]{xcolor}
\usepackage[colorlinks=true,bookmarks=false,linkcolor=NavyBlue,urlcolor=NavyBlue,citecolor=NavyBlue,breaklinks]{hyperref}
\usepackage{amsmath}
\usepackage{amsfonts}
\usepackage{tabularx}
\usepackage{graphicx}
\usepackage{times}
\usepackage{mathtools}
\usepackage{braket}
\usepackage{enumitem}

\newcommand{\real}{\operatorname{Re}}

\newcommand{\parti}[2]{\frac{\partial #1}{\partial #2}}

\newcommand{\abs}[1]{\left|#1\right|}
\newcommand{\bk}[1]{\left(#1\right)}
\newcommand{\Bk}[1]{\left[#1\right]}

\newcommand{\trace}{\operatorname{tr}}

\newcommand{\floor}[1]{\left\lfloor#1\right\rfloor}
\newcommand{\ceil}[1]{\left\lceil#1\right\rceil}

\begin{document}

\title{Quantum limit to subdiffraction incoherent optical imaging}

\author{Mankei Tsang}
\email{mankei@nus.edu.sg}
\homepage{http://mankei.tsang.googlepages.com/}
\affiliation{Department of Electrical and Computer Engineering,
  National University of Singapore, 4 Engineering Drive 3, Singapore
  117583}

\affiliation{Department of Physics, National University of Singapore,
  2 Science Drive 3, Singapore 117551}

\date{\today}


\begin{abstract}
  The application of quantum estimation theory to the problem of
  imaging two incoherent point sources has recently led to new
  insights and better measurements for incoherent imaging and
  spectroscopy.  To establish a more general limit beyond the case of
  two sources, here I evaluate a quantum bound on the Fisher
  information that can be extracted by any far-field optical
  measurement about the moments of a subdiffraction object. The bound
  matches the performance of a spatial-mode-demultiplexing (SPADE)
  measurement scheme in terms of its scaling with the object size,
  indicating that SPADE is close to quantum-optimal. Coincidentally,
  the result is also applicable to the estimation of diffusion
  parameters with a quantum probe subject to random displacements.
\end{abstract}

\maketitle

\section{Introduction}
The fundamental resolution of optical imaging can be framed as a
problem of quantum estimation \cite{helstrom}: With any measurement
permitted by quantum mechanics, how well can one estimate unknown
parameters from the light? While Helstrom laid the foundation of
quantum estimation theory and first applied it to incoherent imaging
\cite{helstrom}, it was not until recently that this approach yielded
genuine surprises on the age-old topic.  Through the computation of
the quantum Fisher information (QFI), it was found that the separation
between two sub-Rayleigh incoherent point sources can be estimated
much more accurately than previously realized \cite{tnl}.  This
discovery has since led to new insights and better measurements for
incoherent imaging and spectroscopy
\cite{tnl,tsang16c,tsang_semiclassical,tsang16,sliver,tnl2,nair_tsang16,lupo,ant,krovi16,lu18,rehacek16,pearce17,yang17,kerviche17,chrostowski17,rehacek17,rehacek17a,backlund18,zhou18v2,napoli18,yu18,prasad18,larson18,tsang_comment18}.
Experimental demonstrations have also been reported
\cite{tham16,tang16,yang16,paur16,parniak18,donohue18,paur18,hassett18,zhou18a}.

Generalizing such results for arbitrary source distributions is much
more difficult, as the quantum state may depend on infinitely many
spatial modes and infinitely many parameters. Some progress has been
made in Refs.~\cite{tsang16c,tsang_semiclassical}, which evaluate the
performance of a spatial-mode-demultiplexing (SPADE) measurement for
estimating the moments of any subdiffraction
object. Reference~\cite{tsang16c} also proves quantum bounds for
location and scale parameters and conjectures that SPADE may be
quantum-optimal for general imaging. A similar conjecture was raised
earlier by Krovi, Guha, and Shapiro in Ref.~\cite{krovi16}.  Zhou and
Jiang have recently taken a major step towards proving the conjectures
\cite{zhou18v2}: Using novel arguments that do not resort to the QFI,
they propose limits on the scaling of the Fisher information with
respect to the object size for any moment parameter.  Their bounds may
have issues concerning their precise values and validity, however, as
elaborated in Appendix~\ref{sec_zhou}.

To derive a limit using more standard quantum estimation theory, here
I evaluate an upper bound on the QFI for the moment-estimation
problem. The result matches the performance of SPADE evaluated in
Refs.~\cite{tsang16c,tsang_semiclassical} in terms of the object-size
scaling, indicating that SPADE is close to quantum-optimal. While the
end result looks similar to those of Zhou and Jiang, the use of the
QFI here leads to a bound that overcomes the issues in
Ref.~\cite{zhou18v2} and sets a more absolute and computable quantum
limit for incoherent imaging.

\section{\label{sec_background} Quantum optics and quantum estimation theory}
\begin{figure}[htbp!]
  \centerline{\includegraphics[width=0.45\textwidth]{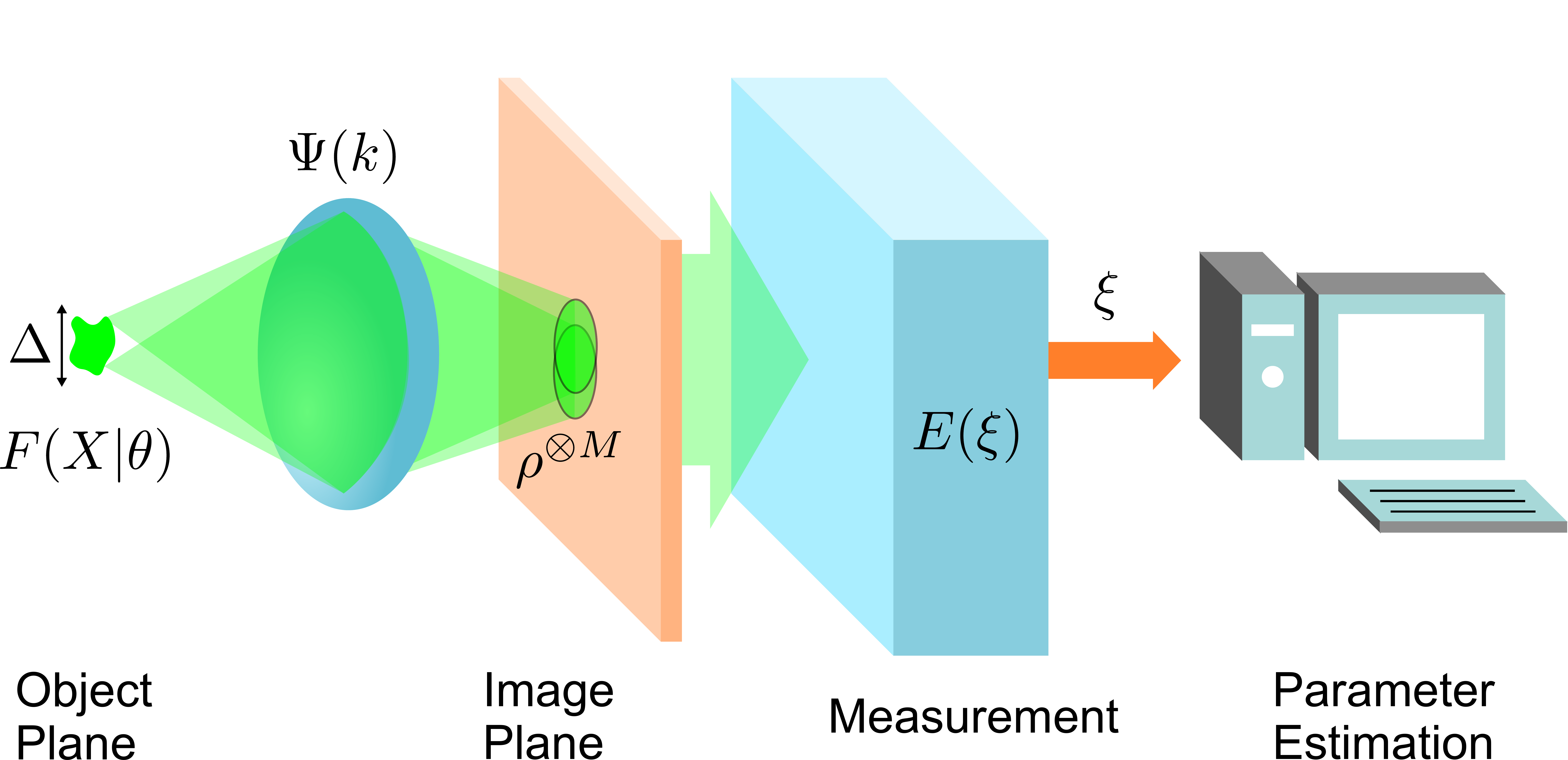}}
  \caption{\label{imaging_figure}A far-field incoherent optical
    imaging system. $F(X|\theta)$ is the object intensity
    distribution, $\Delta$ is its characteristic width, $\Psi(k)$ is
    the optical transfer function (OTF) of the imaging system,
    $\rho^{\otimes M}$ is the quantum state of light in $M$ temporal
    modes on the image plane, $E$ is the positive operator-valued
    measure (POVM) that models the measurement, and $\xi$ is a
    measurement outcome.}
\end{figure}

Consider the far-field imaging of quasi-monochromatic incoherent
optical sources, as depicted by Fig.~\ref{imaging_figure}. The quantum
state of light in $M$ temporal modes can be modeled as the tensor
product $\rho^{\otimes M}$, where
\begin{align}
\rho &= (1-\epsilon)\rho_0 + \epsilon \rho_1,
\label{rho}
\end{align}
$\epsilon\ll 1$ is the expected photon number per temporal mode,
$\rho_0$ is the vacuum state, $\rho_1$ is the one-photon state, and
$O(\epsilon^2)$ terms are ignored \cite{gottesman,stellar,tnl}.
Assuming scalar paraxial waves \cite{goodman} and the imaging of
sources in one transverse dimension for simplicity, the one-photon
state on the image plane is given by \cite{tnl,tsang16c}
\begin{align}
\rho_1 &= \int dX F(X|\theta) e^{-i\hat kX}\ket{\psi}\bra{\psi} e^{i\hat kX},
\label{rho1}
\\
\ket{\psi} &= \int dk \Psi(k)\ket{k},
\end{align}
where $F(X|\theta)$ is the normalized object intensity distribution
with $\int dX F(X|\theta) = 1$, $X$ is the object-plane coordinate,
$\theta = (\theta_1,\theta_2,\dots)$ is a vector of unknown
parameters, $\hat k$ is the one-photon spatial-frequency (momentum)
operator, $\ket{k}$ is the one-photon eigenket that satisfies
$\hat k\ket{k} = k\ket{k}$ and $\braket{k|k'} = \delta(k-k')$, and
$\Psi(k)$ is the optical transfer function (OTF) of the imaging system
\cite{goodman}.  The diffraction limit introduces a finite bandwidth
to $\Psi(k)$, and the Fourier transform of $\Psi(k)$ gives the
point-spread function.  $X$ and $\hat k$ are normalized with respect
to the magnification factor and the OTF bandwidth such that they are
unitless. While this work will focus on imaging, note that
Eq.~(\ref{rho1}) also describes a quantum object in initial state
$\ket{\psi}$ subject to random displacements with unknown statistics
\cite{hall09,vidrighin,ng16}.

Any measurement can be modeled by a positive operator-valued measure
(POVM) $E$ \cite{helstrom,hayashi}, such that the probability of a
measurement outcome $\xi$ conditioned on $\theta$ is
\begin{align}
  P(\xi|\theta) = \trace E(\xi)\rho^{\otimes M},
\end{align}
where $\trace$ denotes the operator trace. If the measurement consists
of passive linear optics and photon counting, the standard Poisson
model in optical astronomy and fluorescence microscopy
\cite{goodman_stat,feigelson,zmuidzinas03,ram,deschout,small14,chao16,diezmann17,tnl2,tsang_semiclassical}
is retrieved in the ``ultraviolet'' limit of $\epsilon \to 0$ and
$M\to\infty$, with $N \equiv M\epsilon$, the expected photon number in
all modes, held constant \cite{tnl}.

Denoting the partial derivative with respect to $\theta_\mu$ by the
comma notation $P_{,\mu} \equiv \partial P/\partial \theta_\mu$, the
Fisher information matrix is given by
\begin{align}
J_{\mu\nu}(P) &\equiv \sum_{\xi}
\frac{P_{,\mu}(\xi|\theta)P_{,\nu}(\xi|\theta)}
{P(\xi|\theta)},
\label{J}
\end{align}
which plays a fundamental role in parameter estimation theory and can
be used to set Cram\'er-Rao lower error bounds
\cite{vantrees,tsang16}. In the context of imaging, the Fisher
information has been proposed by many as the fundamental measure of
resolution
\cite{farrell66,helstrom70b,tsai79,bettens,vanaert,ram,deschout,small14,chao16,diezmann17}.
In recent years, it has become especially popular in fluorescence
microscopy \cite{ram,deschout,small14,chao16,diezmann17}.

In quantum estimation theory, it is known \cite{helstrom,hayashi,tnl}
that, for any POVM,
\begin{align}
  J(P) &\le N K(\rho_1),
\label{qcrb}
\\
K_{\mu\nu}(\rho_1) &= \trace \rho_{1,\mu} L_\nu,
\quad
\rho_{1,\mu} =\frac{1}{2}\bk{L_\mu \rho_1 + \rho_1 L_\mu},
\label{qfi}
\end{align}
where the matrix inequality means that $NK-J$ is
positive-semidefinite. Appendix~\ref{sec_thermal} proves that
Eq.~(\ref{qcrb}) in fact holds for any thermal state with arbitrary
$\epsilon$. The QFI matrix $NK(\rho_1)$ thus serves as an even more
fundamental measure of resolution that depends only on the quantum
state and holds for any measurement.

\section{Quantum bound based on an alternative Choi-Kraus representation}
\begin{figure}[htbp!]
\centerline{\includegraphics[width=0.45\textwidth]{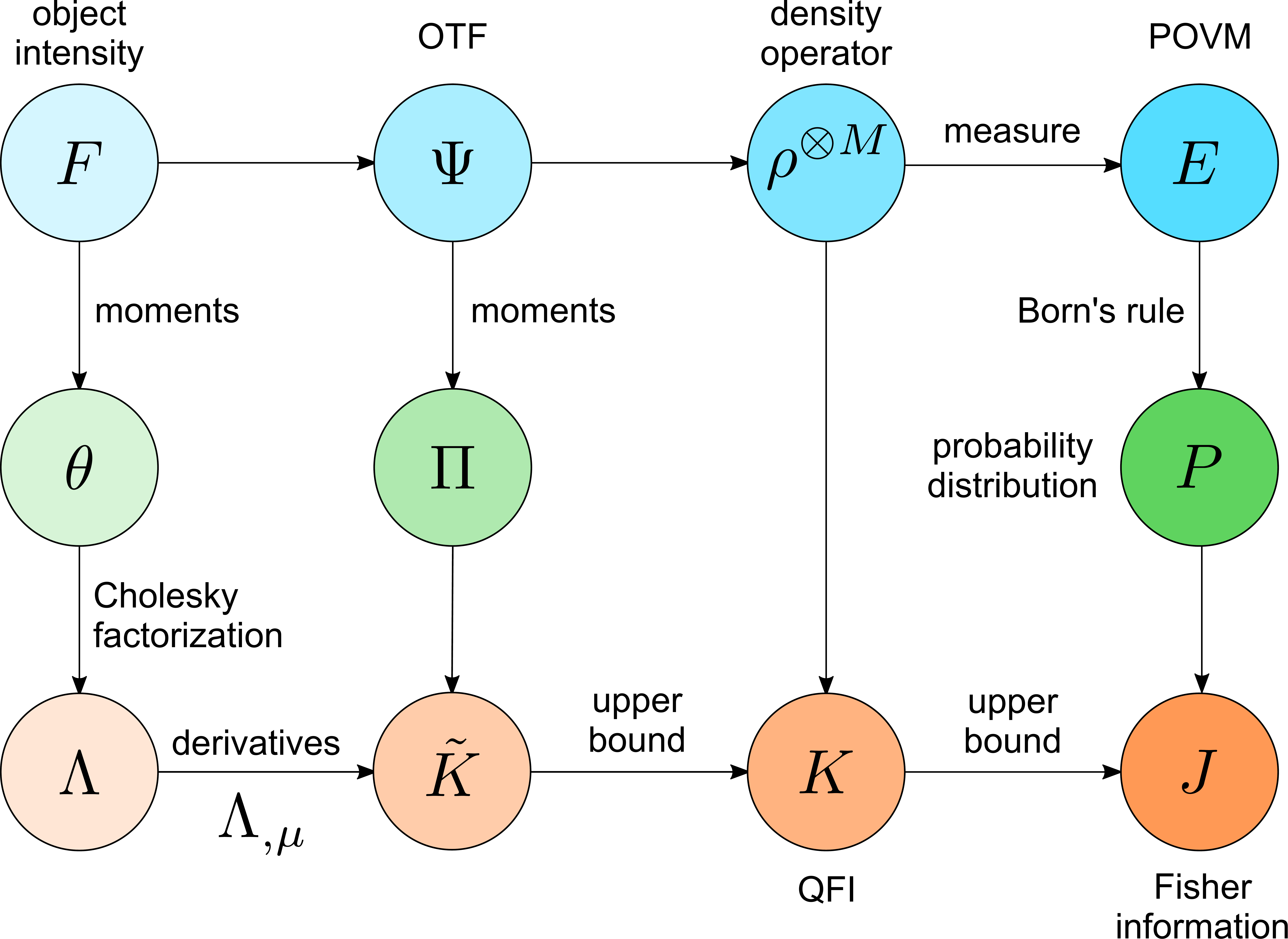}}
\caption{\label{flowchart}A summary of the relationships among the key
  quantities in this work.}
\end{figure}

Define the object moment parameters as
\begin{align}
\theta_{\mu} &\equiv \int dX F(X|\theta) X^\mu,
\quad \mu \in \mathbb N,
\label{moments}
\end{align}
with $\theta_0 = 1$. Under benign conditions, each moment sequence
determines $F$ uniquely \cite{dunkl}, so there is little loss of
generality by parametrizing the imaging problem in terms of the
moments.  Expanding $\exp(-i\hat kX)$ in the Taylor series
$\sum_{q=0}^\infty (-i\hat k)^q X^q/q!$, I can rewrite
Eq.~(\ref{rho1}) as
\begin{align}
\rho_1 &= \sum_{q=0}^\infty \sum_{p=0}^\infty
\theta_{q+p}  \frac{(-i\hat k)^q}{q!}\ket{\psi}\bra{\psi}
\frac{(i\hat k)^p}{p!}.
\label{expand}
\end{align}
Assume that the support of $F(X|\theta)$ has an infinite number of
points, such that $\int dX F(X|\theta) \mathcal P^2(X) > 0$ for any
nonzero polynomial $\mathcal P$, and the Hankel matrix $\theta_{q+p}$
is positive-definite \cite{dunkl}. The Cholesky factorization can then
be used to write
\begin{align}
\theta_{q+p} &= \sum_{r=0}^\infty \Lambda_{qr}\Lambda_{pr},
\label{cholesky}
\end{align}
where $\Lambda$ is a real lower-triangular matrix with strictly
positive diagonal elements \cite{horn}.  Equation~(\ref{expand})
becomes
\begin{align}
\rho_1 &= \sum_{r=0}^\infty A_r \ket{\psi}\bra{\psi} A_r^\dagger,
&
A_r &\equiv \sum_{q=0}^\infty \Lambda_{qr}\frac{(-i\hat k)^q}{q!},
\end{align}
where $\{A_r\}$ are Choi-Kraus operators \cite{hayashi} and $\dagger$
denotes the Hermitian conjugate.  It can be shown via purification
\cite{escher} that an upper bound on the QFI is
\begin{align}
K(\rho_1) &\le \tilde K, 
\quad
\tilde K_{\mu\nu} =
4\real\bk{B_\mu B_\nu+C_{\mu\nu}},
\\
B_\mu &\equiv \sum_{r=0}^\infty \bra{\psi}A_r^\dagger A_{r,\mu}\ket{\psi},
\\
C_{\mu\nu}&\equiv 
\sum_{r=0}^\infty \bra{\psi}A_{r,\nu}^\dagger A_{r,\mu}\ket{\psi}.
\end{align}
Defining the positive-semidefinite matrix
\begin{align}
  \Pi_{pq} &\equiv \frac{1}{p!q!}
  \bra{\psi}(i\hat k)^p(-i\hat k)^q\ket{\psi}
=\frac{i^{p-q}}{p!q!}\int dk |\Psi(k)|^2 k^{p+q},
\label{Pi}
\end{align}
which consists of the OTF moments, I obtain
\begin{align}
B_\mu &= \trace \Pi \Lambda \Lambda_{,\mu}^\top,
&
C_{\mu\nu} &= \trace \Pi \Lambda_{,\mu} \Lambda_{,\nu}^\top,
\end{align}
where $\top$ denotes the transpose. Assume that the OTF magnitude is
even, viz., $|\Psi(k)|^2 = |\Psi(-k)|^2$, such that $\Pi$ is real and
symmetric ($\Pi = \Pi^\top$), and $B_\mu$ and $C_{\mu\nu}$ are also
real.  To evaluate $B_\mu$, first note that
\begin{align}
B_\mu &= \trace \Pi \Lambda \Lambda_{,\mu}^\top
= \trace (\Pi \Lambda \Lambda_{,\mu}^\top)^\top
= \trace \Lambda_{,\mu} \Lambda^\top \Pi^\top
\nonumber\\&
=\trace \Pi^\top \Lambda_{,\mu} \Lambda^\top 
=\trace \Pi \Lambda_{,\mu} \Lambda^\top.
\end{align}
Then the normalization of $\rho_1$ can be used to show
\begin{align}
\trace \rho_1 &=  \trace \Pi\Lambda\Lambda^\top = 1,
\\
\trace \rho_{1,\mu} &= \trace \Pi \Lambda_{,\mu}\Lambda^\top
+\trace \Pi \Lambda \Lambda_{,\mu}^\top = 2B_\mu = 0.
\end{align}
Hence 
\begin{align}
\tilde K_{\mu\nu} &= 4C_{\mu\nu} = 
4 \trace \Pi\Lambda_{,\mu}\Lambda_{,\nu}^\top.
\label{Ktilde}
\end{align}
Figure~\ref{flowchart} summarizes the relationships among the key
quantities in this work. 

As the right-hand side of Eq.~(\ref{Ktilde}) consists of infinite
sums, their convergence is needed for $\tilde K$ to be a nontrivial
upper bound on the QFI. Appendix~\ref{sec_cond} proves that
$|\tilde K_{\mu\nu}| < \infty$ if $|\Psi(k)|^2$ is bandlimited or
Gaussian ($\propto \exp[-k^2/(2\beta^2)]$) and $F(X|\theta)$ is any
probability density with compact support in the Szeg\H{o} class
\cite{berg11} or Gaussian ($\propto \exp[-X^2/(2\Delta^2)]$). If both
are Gaussian, a further condition is $\beta\Delta < 1/2$. These are
sufficient conditions but already quite general; $\tilde K$ may
converge under more relaxed conditions.

\section{Quantum bound in the subdiffraction regime}
Although the QFI and its upper bound $\tilde K$ are functions of
infinitely many parameters in general, the goal of this work is to
show that $\tilde K_{\mu\mu}$ obeys a universal behavior when the
parameters correspond to a subdiffraction regime.  Let $\Delta > 0$ be
a characteristic width of $F(X|\theta)$ around $X = 0$, such that
$\theta_\mu = O(\Delta^\mu)$, where the big O notation denotes terms
on the order of the argument and is defined by
\begin{align}
\lim_{\Delta\to 0} \abs{\frac{O[f(\Delta)]}{f(\Delta)}} < \infty.
\label{bigO}
\end{align}
Recall that $X$ has been normalized with respect to the magnification
factor and OTF bandwidth; the subdiffraction regime can therefore be
defined by $\Delta \ll 1$ \cite{tsang16c,tsang_semiclassical}.

The dependence of the Cholesky factor $\Lambda$ on $\theta$ can be
studied via the recursive relation \cite{gentle17}
\begin{align}
\Lambda_{qr} &= 
\begin{cases}
\sqrt{\theta_{2q}-\sum_{s=0}^{q-1}(\Lambda_{qs})^2},
&
r = q,
\\
\bk{\theta_{q+r}-\sum_{s=0}^{r-1} \Lambda_{qs}\Lambda_{rs}}/\Lambda_{rr},
&
r < q,
\\
0,
&
r > q,
\end{cases}
\label{recursive}
\end{align}
starting from $\Lambda_{00} = \sqrt{\theta_0} = 1$.
Equation~(\ref{recursive}) can be differentiated to give
\begin{widetext}
\begin{align}
\Lambda_{qr,\mu} &= 
\begin{cases}
\bk{\delta_{\mu}^{2q}-2\sum_{s=0}^{q-1}\Lambda_{qs}\Lambda_{qs,\mu}}/(2\Lambda_{qq}),
& r = q,\\
\Bk{\delta_\mu^{q+r}-\sum_{s=0}^{r-1}
(\Lambda_{qs,\mu}\Lambda_{rs} + \Lambda_{qs}\Lambda_{rs,\mu})
-\Lambda_{qr}\Lambda_{rr,\mu}}/\Lambda_{rr},
& r < q,\\
0, & r > q,
\end{cases}
\label{dLambda}
\end{align}
\end{widetext}
where $\delta^a_b$ is the Kronecker delta. Since the diagonal elements
of the Cholesky factor $\Lambda$ are all strictly positive, all
$\Lambda$ and $\Lambda_{,\mu}$ elements are finite, and a dimensional
analysis of Eqs.~(\ref{recursive}) and (\ref{dLambda}) gives
\begin{align}
  \Lambda_{qr} &= O(\Delta^{q}),
  \label{dim1}
\\
  \Lambda_{qr,\mu} &= O(\Delta^{q-\mu}).
  \label{dim2}
\end{align}
Inspecting the dependence of the $\Lambda$ elements on a given
$\theta_\mu$ with even $\mu$, starting from the upper-left corner and
going row by row, one can see that the dependence does not appear
until the diagonal element $\Lambda_{qq}$ with $q = \mu/2$. In other
words,
\begin{align}
\Lambda_{qr,\mu} &= 
\begin{cases}
0, & q \le \mu/2, r < \mu/2,
\\
1/(2\Lambda_{qq}) = O(\Delta^{-\mu/2}), & q = r = \mu/2,
\\
o(\Delta^{-\mu/2}), & q > \mu/2,
\end{cases}
\label{DLeven}
\end{align}
where the small o notation denotes terms that are asymptotically
negligible relative to the argument and is defined by
\begin{align}
\lim_{\Delta \to 0} \abs{\frac{o(f(\Delta))}{f(\Delta)}} = 0.
\end{align}
Thus only one element in $\Lambda_{,\mu}$ is $O(\Delta^{-\mu/2})$, and
the rest of the elements are all in higher orders. I can then express
Eq.~(\ref{Ktilde}) as
\begin{align}
\tilde K_{\mu\mu} &= 4\sum_{s,t}
\Pi_{st}\sum_r \Lambda_{tr,\mu}\Lambda_{sr,\mu}
\\
&= 4\Pi_{qq} \bk{\Lambda_{qq,\mu}}^2 + 
\nonumber\\&\quad
4\sum_{s+t>\mu} \Pi_{st} \sum_r \Lambda_{tr,\mu}\Lambda_{sr,\mu},
\quad q = \frac{\mu}{2}.
\label{Kmumu}
\end{align}
Recall that $\hat k$ has been normalized with respect to the OTF
width, and usual OTFs, such as bandlimited and Gaussian functions,
have finite moments. Thus $\Pi_{st} = O(1)$, only
$4\Pi_{qq} (\Lambda_{qq,\mu})^2$ is $O(\Delta^{-\mu})$, and the rest
of the terms on the right-hand side of Eq.~(\ref{Kmumu}) are all
$o(\Delta^{-\mu})$.  Assuming $\tilde K_{\mu\mu} < \infty$, the
infinite sum in Eq.~(\ref{Kmumu}) converges to $o(\Delta^{-\mu})$, and
$\tilde K_{\mu\mu}$ can be approximated as
\begin{align}
  \tilde K_{\mu\mu} & =O(\Delta^{-\mu})
\approx 4\Pi_{qq}\bk{\Lambda_{qq,\mu}}^2
  = \frac{\bra{\psi}\hat k^{2q}\ket{\psi}}{q!^2(\Lambda_{qq})^2},
\quad
q = \frac{\mu}{2}.
\label{Keven}
\end{align}

If $\mu$ is odd, the dependence of $\Lambda$ on a given $\theta_\mu$
starts to appear only on the row $q = (\mu+1)/2$ in the elements
$\Lambda_{q\,q-1}$ and $\Lambda_{qq}$. Specifically,
\begin{widetext}
\begin{align}
\Lambda_{qr,\mu} &= 
\begin{cases}
0, & q \le (\mu+1)/2, r < (\mu-1)/2,
\\
1/(\Lambda_{rr}) = O(\Delta^{-(\mu-1)/2}), & q =(\mu+1)/2, r = (\mu-1)/2,
\\
-\Lambda_{q\,q-1}/(\Lambda_{qq}\Lambda_{q-1\,q-1})
=O(\Delta^{-(\mu-1)/2}), & q = r = (\mu+1)/2,
\\
o(\Delta^{-(\mu-1)/2}), & q > (\mu+1)/2.
\end{cases}
\label{DLodd}
\end{align}
\end{widetext}
Now there are two $O(\Delta^{-(\mu-1)/2})$ leading-order terms in
$\Lambda_{qr,\mu}$. I can express Eq.~(\ref{Ktilde}) as
\begin{align}
\tilde K_{\mu\mu} &= 4\Pi_{qq}\Bk{\bk{\Lambda_{q\,q-1,\mu}}^2+\bk{\Lambda_{qq,\mu}}^2} + 
\nonumber\\&\quad
4 \sum_{s+t>\mu+1} \Pi_{st}\sum_r \Lambda_{tr,\mu}\Lambda_{sr,\mu},
\quad
q = \frac{\mu+1}{2},
\label{Kmumu_odd}
\end{align}
where
$4\Pi_{qq}[(\Lambda_{q\,q-1,\mu})^2+(\Lambda_{qq,\mu})^2]=O(\Delta^{-(\mu-1)})$
and the rest of the terms are all $o(\Delta^{-(\mu-1)})$. Assuming
again $\tilde K_{\mu\mu} < \infty$, I obtain
\begin{align}
\tilde K_{\mu\mu} &=O(\Delta^{-(\mu-1)})
\approx 4\Pi_{qq}\Bk{\bk{\Lambda_{q\,q-1,\mu}}^2+\bk{\Lambda_{qq,\mu}}^2}
\nonumber\\&= \frac{4\bra{\psi}\hat k^{2q}\ket{\psi}}
{q!^2(\Lambda_{q-1\,q-1})^2}
\Bk{1 + \bk{\frac{\Lambda_{q\,q-1}}{\Lambda_{qq}}}^2},
\quad q = \frac{\mu+1}{2}.
\label{Kodd}
\end{align}
Equation~(\ref{Keven}) for even $\mu$ and Eq.~(\ref{Kodd}) for odd
$\mu$ can be summarized as
\begin{align}
J_{\mu\mu}(P) \le
N K_{\mu\mu}(\rho_1)
\le
N\tilde K_{\mu\mu} = N O(\Delta^{-2\floor{\mu/2}}),
\label{bound}
\end{align}
which sets a lower bound on the mean-square error $\textrm{MSE}_\mu$
of any unbiased estimator of a moment $\theta_\mu$ via the
Cram\'er-Rao bound
$\textrm{MSE}_\mu \ge (J^{-1})_{\mu\mu} \ge 1/J_{\mu\mu}$
\cite{vantrees}.

\section{Discussion}
Equations~(\ref{Keven}), (\ref{Kodd}), and (\ref{bound}) are the
central results of this work.  The scaling of Eq.~(\ref{bound}) with
respect to $\Delta$ matches the performance of SPADE for moment
estimation evaluated in Refs.~\cite{tsang16c,tsang_semiclassical}.
The Fisher information for direct imaging is $J_{\mu\mu} = N O(1)$ in
the subdiffraction regime, so substantial improvements can be obtained
for $\mu \ge 2$ \cite{tsang16c,tsang_semiclassical}. For $\mu = 1,2$,
the inverse of Eq.~(\ref{bound}) also matches an
$O(\Delta^{2\mu-2})/N$ quantum error bound computed in
Appendix~\ref{sec_convexity} via the convexity of QFI.

For a more sobering perspective, consider the signal-to-noise ratio
(SNR), defined here as $\theta_\mu^2 = O(\Delta^{2\mu})$ divided by
the mean-square error. Equation~(\ref{bound}) then suggests that a
quantum limit on the SNR is
\begin{align}
\textrm{QSNR}_\mu &\equiv N \tilde K_{\mu\mu} \theta_\mu^2=
  NO(\Delta^{2\lceil\mu/2\rceil}).
  \label{snr}
\end{align}
While it remains a significant improvement over the
$NO(\Delta^{2\mu})$ SNR for direct imaging, Eq.~(\ref{snr}) still
decreases for smaller $\Delta$, especially for higher moments, and
decays in a roughly exponential fashion with increasing $\mu$ for a
given $\Delta$ in the subdiffraction regime, as shown more carefully
in Appendix~\ref{sec_prefactor}.  This difficulty with higher moments
is known in the context of SPADE \cite{tsang16c,tsang_semiclassical},
but the quantum limit here proves that it is fundamental for any
measurement.

Although Eq.~(\ref{bound}) assumes one-dimensional imaging, previous
studies of two-dimensional imaging in quantum estimation theory
\cite{ant,tsang16c,tsang_semiclassical,zhou18v2} show no new surprises,
and it is reasonable to conjecture that the quantum limit on the
Fisher information becomes $N O(\Delta^{-2\floor{|\mu|/2}})$---the
same as the SPADE performance---where $|\mu| = \sum_j\mu_j$ is the
total moment order \cite{tsang16c,tsang_semiclassical}.

Unlike Zhou and Jiang's Theorem 1 in Ref.~\cite{zhou18v2}, the quantum
bound here does not depend on the POVM and is more amenable to
approximation or numerical computation. The scaling of
Eq.~(\ref{bound}) with $\Delta$ for odd moments is also tighter than
that suggested by their Theorem 1. Furthermore, their Theorem 3 makes
a questionable assumption about the optimal POVM.
Appendix~\ref{sec_zhou} presents a review of Ref.~\cite{zhou18v2} and
highlights these issues. The use of the QFI here, on the other hand,
guarantees that Eq.~(\ref{bound}) holds for any POVM.

Beyond imaging, Eq.~(\ref{rho1}) also describes a quantum object
subject to random displacements with unknown and possibly non-Gaussian
statistics.  $\Delta$ is then a measure of the displacement magnitude.
The result here can therefore be applied to the estimation of
diffusion parameters with a quantum probe in the weak-signal
($\Delta \ll 1$) regime, without the need to assume Gaussian
statistics as in prior works \cite{hall09,vidrighin,ng16}.  Potential
applications include magnetometry \cite{hall09}, optical
interferometry \cite{vidrighin}, and optomechanical force sensing
\cite{ng16}.

\section{Conclusion}
I have proposed a general quantum limit to subdiffraction incoherent
imaging in terms of moment estimation, going far beyond the simple
example of two point sources in previous studies. This limit does not
depend on the measurement and is also tight in terms of its scaling
with the object size, thus setting a fundamental criterion of
resolution for far-field incoherent imaging, with prime applications
being observational astronomy and fluorescence microscopy.

Looking forward, many open problems still remain, including a more
precise evaluation of the QFI, a more detailed comparison with SPADE,
generalizations for more dimensions and other types of sources,
derivations of tighter multiparameter quantum bounds, and an
experimental demonstration of quantum-limited measurements for more
general objects. As the light sources are classical and the
measurements require only far-field linear optics and photon counting
\cite{tnl2,tsang16c,tsang_semiclassical}, a clear path towards
practical applications of the quantum-inspired technology can be
envisioned, with the quantum limit serving as the ultimate yardstick.

\section*{Acknowledgments}
This work was supported by the Singapore Ministry of Education
Academic Research Fund Tier 1 Project R-263-000-C06-112.

\appendix

\section{\label{sec_thermal}Quantum bounds for thermal states}

\subsection{A bound on the QFI}
Let $\{a_j\}$ be the bosonic annihilation operators with respect to a
set of optical modes and
\begin{align}
\rho = \int D\alpha \Phi(\alpha) \ket{\alpha}\bra{\alpha},
\end{align}
where $\alpha \equiv (\alpha_1,\alpha_2,\dots)^\top$ is a column
vector of complex amplitudes, $D\alpha \equiv \prod_j d^2\alpha_j$,
$\Phi$ is the Glauber-Sudarshan distribution, and $\ket{\alpha}$ is a
coherent state that satisfies $a_j\ket{\alpha} = \alpha_j\ket{\alpha}$
\cite{helstrom}. For a thermal state,
\begin{align}
\Phi &= \frac{1}{\det(\pi\Gamma)}\exp\bk{-\alpha^\dagger \Gamma^{-1}\alpha},
\end{align}
where $\Gamma > 0$ is the mutual coherence matrix.  Helstrom has shown
in Sec.~V of Ref.~\cite{helstrom68} (see also Sec.~VIII 6(b) of
Ref.~\cite{helstrom}) that the QFI is
\begin{align}
K_{\mu\nu}(\rho) &=  \trace \Gamma_{,\mu}\Upsilon_\nu,
\label{Kthermal}
\end{align}
where $\Upsilon_\mu$ is a Hermitian matrix that satisfies
\begin{align}
\Gamma_{,\mu} &= \frac{1}{2}
\Bk{(I+\Gamma)\Upsilon_\mu\Gamma+ \Gamma\Upsilon_\mu(I+\Gamma)},
\label{Upsilon}
\end{align}
and $I$ is the identity matrix. The QFI is an upper bound on the
Fisher information for any POVM \cite{hayashi}, viz.,
\begin{align}
J(P) &\le K(\rho^{\otimes M}) = M K(\rho).
\label{qfibound}
\end{align}

To obtain a simpler bound than
Eqs.~(\ref{Kthermal})--(\ref{qfibound}), diagonalize $\Gamma$ in terms
of its eigenvalues $\{\gamma_j\}$ and orthonormal eigenvectors
$\{e_j\}$ as
\begin{align}
\Gamma &= \sum_j \gamma_j e_j e_j^\dagger,
\end{align}
where $\{e_j\}$ includes vectors that support $\{\Gamma_{,\mu}\}$ and
$\gamma_j \ge 0$. In terms of this basis, Eqs.~(\ref{Kthermal}) and
(\ref{Upsilon}) can be expressed as \cite{helstrom68}
\begin{align}
K_{\mu\nu}(\rho) &= 
\sum_{j,l}\frac{2 (e_j^\dagger \Gamma_{,\mu} e_l)(e_l^\dagger \Gamma_{,\nu} e_j)}
{\gamma_j+\gamma_l+2\gamma_j\gamma_l}.
\label{Ksum}
\end{align}
Let $u$ be an arbitrary real vector and
$\Gamma' \equiv \sum_\mu u_\mu \Gamma_{,\mu}$.  Since $\Gamma_{,\mu}$
and therefore $\Gamma'$ are Hermitian,
\begin{align}
\sum_{\mu,\nu} u_\mu K_{\mu\nu}(\rho) u_\nu  &= 
\sum_{j,l}\frac{2 |e_j^\dagger \Gamma' e_l|^2}
{\gamma_j+\gamma_l+2\gamma_j\gamma_l}
\le \sum_{j,l}\frac{2 |e_j^\dagger \Gamma' e_l|^2}{\gamma_j+\gamma_l}
\nonumber\\
&= \epsilon \sum_{\mu,\nu} u_\mu K_{\mu\nu}(\Gamma) u_\nu,
\label{sldbound}
\end{align}
where I have extended the definition of the QFI for any
positive-definite matrix as
\begin{align}
K_{\mu\nu}(\Gamma) &= \frac{\trace \Gamma_{,\mu}L_\nu^{(\Gamma)}}{\trace\Gamma},
\label{KGamma}
\\
\Gamma_{,\mu} &= \frac{1}{2}\bk{L_\mu^{(\Gamma)} \Gamma+ \Gamma L_\mu^{(\Gamma)}},
\label{sldGamma}
\end{align}
and $L_\mu^{(\Gamma)}$ is a symmetric logarithmic derivative (SLD) of
$\Gamma$.  Equation~(\ref{sldbound}) results in
\begin{align}
K(\rho) &\le \epsilon K(\Gamma),
&
MK(\rho) &\le NK(\Gamma),
\end{align}
which can be combined with Eq.~(\ref{qfibound}) to give
\begin{align}
J(P) &\le K(\rho^{\otimes M}) \le N K(\Gamma).
\label{uvbound}
\end{align}
In other words, rather than computing $K(\rho)$ via
Eqs.~(\ref{Kthermal}) and (\ref{Upsilon}), one can compute a looser
quantum bound given by Eqs.~(\ref{KGamma}) and (\ref{sldGamma}) in
terms of the SLDs of $\Gamma$.

\subsection{Ultraviolet and infrared limits}
Let
\begin{align}
\Gamma &= \epsilon g, &\epsilon &= \trace\Gamma, & \trace g &= 1.
\end{align}
In the limit $\epsilon \to 0$, $I+\Gamma \to I$, and the
$\Upsilon_\mu$ defined by Eq.~(\ref{Upsilon}) becomes identical
to the $L_\mu^{(\Gamma)}$ defined by
Eq.~(\ref{sldGamma}). Taking the ultraviolet limit $\epsilon \to 0$
while holding $N \equiv M\epsilon$ constant, I obtain
\begin{align}
\lim_{\epsilon\to 0} M K_{\mu\nu}(\rho) 
&= \lim_{\epsilon\to 0} M\trace \Gamma_{,\mu}\Upsilon_\nu
= N K_{\mu\nu}(\Gamma),
\label{uv}
\end{align}
which means that, in the ultraviolet limit, the QFI approaches
$NK(\Gamma)$, and the second inequality in Eq.~(\ref{uvbound}) becomes
an equality. 

One may also ask what happens in the opposite $\epsilon \to \infty$
``infrared'' limit, which is more applicable to radio and microwave
frequencies or scattered laser sources. Then $I+\Gamma \to \Gamma$,
$\Upsilon_\mu \to \Gamma^{-1}\Gamma_{,\mu} \Gamma^{-1}$, and the
$\epsilon \to\infty$ limit gives
\begin{align}
\lim_{\epsilon\to\infty} K_{\mu\nu}(\rho^{\otimes M}) &=
M \trace \Gamma_{,\mu}\Gamma^{-1}\Gamma_{,\nu}\Gamma^{-1} = MJ(\Phi),
\label{Kstrong}
\end{align}
which is equal to the classical Fisher information with respect to
$\Phi$ \cite{vantrees4}.  Heterodyne detection is sufficient to
achieve this quantum limit, as the Husimi distribution, which governs
the heterodyne statistics, approaches $\Phi$ in the
$\epsilon \to\infty$ limit. For any $\epsilon$, the
classical-simulation technique \cite{demkowicz} can also be used to
show that
\begin{align}
K_{\mu\nu}(\rho^{\otimes M}) &\le MJ(\Phi),
\end{align}
since $\Phi$ is positive.

\subsection{Proof of Eq.~(\ref{qcrb}) for any thermal state}
Now suppose that $\epsilon$ does not depend on $\theta$. The SLDs of
$\Gamma$ become the same as the SLDs of $g$, resulting in
$K(\Gamma) = K(g)$.  Following Refs.~\cite{tnl,tsang16c},
Eq.~(\ref{rho1}) assumes that $g$ is the density matrix of $\rho_1$
with respect to the basis $\{a_j^\dagger\ket{\textrm{vac}}\}$. Since
$K$ is basis-independent, I can write
\begin{align}
K(\Gamma) &= K(g) = K(\rho_1),
\label{Kg}
\end{align}
which can be combined with Eq.~(\ref{uvbound}) to give
Eq.~(\ref{qcrb}). Hence Eq.~(\ref{qcrb}) in fact holds for any thermal
state with arbitrary $\epsilon$. The right-hand side of
Eq.~(\ref{qcrb}) is equal to the QFI for a thermal state in the
ultraviolet limit, as shown by Eq.~(\ref{uv});
Sec.~\ref{sec_background} arrives at the same result by making the
$\epsilon \ll 1$ approximation at the beginning.

Consider the QFI per photon defined as
\begin{align}
\kappa(\epsilon) &\equiv \frac{K(\rho)}{\epsilon},
\\
\kappa_{\mu\nu}(\epsilon) &= 
\sum_{j,l}\frac{2 (e_j^\dagger g_{,\mu} e_l)(e_l^\dagger g_{,\nu} e_j)}
{\lambda_j+\lambda_l+2\epsilon\lambda_j\lambda_l}
\nonumber\\&=
\sum_{j,l}\frac{2 \bra{e_j} \rho_{1,\mu} \ket{e_l}
\bra{e_l} \rho_{1,\nu} \ket{e_j}}
    {\lambda_j+\lambda_l+2\epsilon\lambda_j\lambda_l},
\label{perphoton}
\end{align}
where $\{\lambda_j \equiv \gamma_j/\epsilon\}$ are the eigenvalues of
$g$ and also $\rho_1$ and
$\{\ket{e_j}\equiv\sum_l e_{jl}a_l^\dagger\ket{\textrm{vac}}\}$ are
the eigenkets of $\rho_1$. It is obvious that $\kappa(\epsilon)$ is a
nonincreasing function of $\epsilon$, viz.,
\begin{align}
\kappa(\epsilon') \le \kappa(\epsilon) \textrm{ if }\epsilon' >
\epsilon,
\end{align}
with the supremum achieved at
$\lim_{\epsilon\to 0}\kappa(\epsilon) = K(\rho_1)$. This behavior is
consistent with the explicit calculations of $\kappa(\epsilon)$ in
Refs.~\cite{nair_tsang16,lupo} via other methods.

\section{\label{sec_cond}Sufficient conditions for
  $|\tilde K_{\mu\nu}| < \infty$}

Since the $\Pi$ matrix given by Eq.~(\ref{Pi}) is
positive-semidefinite, the $\tilde K$ matrix given by
Eq.~(\ref{Ktilde}) is Gramian \cite{horn} and also
positive-semidefinite, with
\begin{align}
\tilde K_{\mu\mu} &\ge 0,
&
|\tilde K_{\mu\nu}| &\le \sqrt{\tilde K_{\mu\mu}\tilde K_{\nu\nu}}.
\end{align}
It suffices to prove $\tilde K_{\mu\mu} < \infty$ for any $\mu$.
Let
\begin{align}
\tilde\Pi &\equiv W^\top\Pi W,
&
\tilde\Lambda_{,\mu} &\equiv W^{-1}\Lambda_{,\mu},
\end{align}
where $W$ is a real invertible matrix. Then
\begin{align}
\tilde K_{\mu\mu} &= 4\trace\Pi\Lambda_{,\mu}\Lambda_{,\mu}^\top 
= 4\trace \tilde\Pi \tilde\Lambda_{,\mu}\tilde\Lambda_{,\mu}^\top
\nonumber\\
&\le 4||\tilde\Pi|| \cdot ||\tilde\Lambda_{,\mu}\tilde\Lambda_{,\mu}^\top||_1
= 4||\tilde\Pi|| \cdot ||\tilde\Lambda_{,\mu}||_2^2,
\label{Kmumu2}
\end{align}
where $||\cdot||$ is the operator norm, $||\cdot||_1$ is the trace
norm, and $||\cdot||_2$ is the Hilbert-Schmidt norm \cite{holevo11}.
Thus $\tilde K_{\mu\mu} < \infty$ if 
\begin{enumerate}
\item $\tilde\Pi$ is bounded
($||\tilde\Pi|| < \infty$), and 
\item $\tilde\Lambda_{,\mu}$ is Hilbert-Schmidt
  ($||\tilde\Lambda_{,\mu}||_2 < \infty$).
\end{enumerate}

In the following, I assume
\begin{align}
W_{qp} &= w^q \sqrt{q!}\delta_{p}^q,
\end{align}
where $0 < w < \infty$ is an adjustable constant to make the
convergence conditions more general.

\subsection{\label{sec_bounded} Sufficient conditions for
  $||\tilde\Pi|| \le ||\tilde\Pi||_1 < \infty$}
First I prove that $\tilde\Pi$ is in fact trace-class
($||\tilde\Pi||_1 < \infty$) and must therefore be bounded
($||\tilde\Pi|| \le ||\tilde\Pi||_1 < \infty$) \cite{holevo11} if the
OTF is bandlimited or Gaussian.  In the latter case $w$ should be
chosen appropriately.

Since $\tilde\Pi \ge 0$, it is trace-class if
\begin{align}
||\tilde\Pi||_1 &= \trace \tilde\Pi
= \sum_{q=0}^\infty \frac{w^{2q}}{q!}\int dk |\Psi(k)|^2 k^{2q} < \infty.
\end{align}
Two cases are of interest:
\begin{enumerate}[label=(\roman*)]
\item 
For a bandlimited OTF with support in $[-\beta,\beta]$ and $0 < \beta < \infty$,
\begin{align}
\int dk|\Psi(k)|^2 k^{2q} &\le \beta^{2q},
\label{otfbound1}
\\
\trace \tilde\Pi \le \sum_{q=0}^\infty \frac{(w\beta)^{2q}}{q!}
&= \exp[(w\beta)^2],
\end{align}
which converges for any $w$ and $\beta$.
\item For a Gaussian OTF with
standard deviation $\beta$ \cite{gardiner},
\begin{align}
\int dk |\Psi(k)|^2 k^{2q} &= \frac{(2q)!}{q!2^q}\beta^{2q},
\label{otfbound2}
\\
\trace \tilde\Pi &= \sum_{q=0}^\infty \frac{(2q)!}{q!^22^q}(w\beta)^{2q},
\end{align}
which converges if $w\beta < 1/\sqrt{2}$ according to the ratio test
\cite{arfken}.  Thus I should choose a $w$ that satisfies
\begin{align}
w < \frac{1}{\sqrt{2}\beta}.
\label{wgauss}
\end{align}
\end{enumerate}

\subsection{\label{sec_hs}Sufficient conditions for
  $||\tilde\Lambda_{,\mu}||_2 < \infty$}

Next I prove that $\tilde\Lambda_{,\mu}$ is Hilbert-Schmidt if
$F(X|\theta)$ is any probability density with compact support in the
Szeg\H{o} class \cite{szego36,widom66,berg11} or Gaussian.  In the
latter case, $w$ should also be chosen appropriately.

Noting that $\Lambda_{,\mu}$ is lower-triangular, the Hilbert-Schmidt
norm is given by
\begin{align}
||\tilde\Lambda_{,\mu}||_2^2
&=
\trace \tilde\Lambda_{,\mu}\tilde\Lambda_{,\mu}^\top
= 
\sum_{q=0}^\infty \frac{1}{q! w^{2q}}\sum_{r=0}^q \bk{\Lambda_{qr,\mu}}^2 
\nonumber\\
&=\sum_{q=0}^\infty \frac{\eta_q}{q!w^{2q}},
\label{hs}
\\
\eta_q &\equiv \sum_{r=0}^q \bk{\Lambda_{qr,\mu}}^2.
\label{alpha}
\end{align}
For convenience, I normalize the object-plane coordinate $X$ with
respect to the object characteristic width $0 < \Delta < \infty$ as
$X = x\Delta$, such that
\begin{align}
  \theta_{\mu} &= \int dX F(X|\theta) X^\mu = \phi_\mu \Delta^\mu,
  \label{theta_phi}
\\
\phi_\mu &\equiv \int dx f(x|\theta)x^\mu,
\label{phi}
\\
f(x|\theta) &\equiv \Delta F(x\Delta|\theta),
\label{f}
\end{align}
and $\phi_\mu$ and $f(x|\theta)$ are independent of $\Delta$.
Define the Hankel matrix with respect to $\theta$ as
\begin{align}
  \Theta_{qp} = \theta_{q+p},
  \label{Theta}
\end{align}
and the normalized Hankel matrix as
\begin{align}
  \Xi_{qp} = \phi_{q+p}.
  \label{Xi}
\end{align}
Define also the lower-triangular Cholesky
factors $\Lambda$ and $V$ by
\begin{align}
\Theta &= \Lambda\Lambda^\top,
\\
\Xi &= VV^\top.
\label{cholesky2}
\end{align}
Then the matrices are related by
\begin{align}
\Theta &= D \Xi D,
&
\Lambda &= D V,
&
D_{qp} &\equiv \Delta^q \delta_{p}^q.
\end{align}
In particular,
\begin{align}
  \Lambda_{qr} &= \Delta^q V_{qr} = O(\Delta^q),
  \label{Lambda_V}
\end{align}
which verifies Eq.~(\ref{dim1}).  A formula for $\Lambda_{qr,\mu}$ is
\cite{sarkka}
\begin{align}
\Lambda_{qr,\mu} &= \sum_{s=0}^q \Lambda_{qs} T_{sr} 
\bk{\Lambda_{(q)}^{-1}\Theta_{(q),\mu}\Lambda_{(q)}^{-\top}}_{sr},
\label{sarkka}
\\
T_{sr} &\equiv \begin{cases}0, & s < r,\\
1/2, & s = r,\\
1, & s > r,\end{cases}
\end{align}
where the subscript $(q)$ denotes
the $(q+1)$-by-$(q+1)$ upper-left submatrix, viz.,
\begin{align}
  \Lambda_{(q)rs} &= \Lambda_{rs},
\quad
0 \le r \le q,\ 0 \le s \le q,
\\
\Lambda_{(q)}^{-1} &= (\Lambda_{(q)})^{-1},
\quad
\Lambda_{(q)}^{-\top} = [(\Lambda_{(q)})^{-1}]^\top.
\end{align}
Since
\begin{align}
\Theta_{qr,\mu} &= \delta^{q+r}_\mu,
\end{align}
$\Theta_{(q),\mu} = 0$ if $q < \ceil{\mu/2}$,
and Eq.~(\ref{sarkka}) gives
\begin{align}
\Lambda_{qr,\mu} &= 0 \textrm{ if }q < \ceil{\frac{\mu}{2}},
\label{zero_dL}
\end{align}
which is consistent with Eqs.~(\ref{DLeven}) and (\ref{DLodd}).
Suppressing the subscript $(q)$ for clarity, I can also write
\begin{align}
D^{-1}\Theta_{,\mu}D^{-1} &= \Delta^{-\mu}\Theta_{,\mu},
\\
\Lambda^{-1}\Theta_{,\mu}\Lambda^{-\top} &= 
V^{-1}D^{-1}\Theta_{,\mu}D^{-1}V^{-\top} 
= \Delta^{-\mu} Q,
\\
Q &\equiv V^{-1} \Theta_{,\mu}V^{-\top}.
\label{Q}
\end{align}
Equation~(\ref{sarkka}) becomes
\begin{align}
\Lambda_{qr,\mu} &= \Delta^{q-\mu}\sum_{s=0}^q V_{qs} T_{sr} Q_{sr} 
= O(\Delta^{q-\mu}),
\label{Vmu}
\end{align}
which verifies Eq.~(\ref{dim2}). Applying the Cauchy-Schwartz
inequality to Eq.~(\ref{Vmu}), I obtain
\begin{align}
\bk{\Lambda_{qr,\mu}}^2 
&\le \Delta^{2q-2\mu} \Bk{\sum_{s=0}^q \bk{V_{qs}}^2}\Bk{\sum_{s=0}^q \bk{T_{sr} Q_{sr}}^2}
\nonumber\\
&= \Delta^{2q-2\mu} \phi_{2q}\sum_{s=0}^q \bk{T_{sr} Q_{sr}}^2.
\end{align}
This leads to an upper bound on Eq.~(\ref{alpha}) given by
\begin{align}
\eta_q &\le 
\Delta^{2q-2\mu} \phi_{2q}
\sum_{r=0}^q\sum_{s=0}^q \bk{T_{sr} Q_{sr}}^2.
\end{align}
To simplify the double sum, note that $Q$ as defined by Eq.~(\ref{Q})
is symmetric with $Q_{rs} = Q_{sr}$, so it can be shown that
\cite{chang96}
\begin{align}
||Q||_2^2 &=
\sum_{r=0}^q\sum_{s=0}^q \bk{Q_{sr}}^2
= \sum_{s=0}^q \bk{Q_{ss}}^2 + 
2 \sum_{r=0}^q \sum_{s=r+1}^q \bk{Q_{sr}}^2
\nonumber\\
&\ge 
2 \sum_{r=0}^q\sum_{s=0}^q \bk{T_{sr} Q_{sr}}^2,
\end{align}
leading to
\begin{align}
\eta_q &\le \frac{\Delta^{2q-2\mu} \phi_{2q}}{2}  ||Q||_2^2.
\label{alpha2}
\end{align}
With Eq.~(\ref{Q}), $||Q||_2$ can be bounded as
\begin{align}
||Q||_2 &\le ||V_{(q)}^{-1}||^2 \cdot ||\Theta_{(q),\mu}||_2
\le ||\Xi_{(q)}^{-1}||\sqrt{\mu+1},
\label{Q2}
\end{align}
where I have restored the subscript $(q)$ for emphasis and used the
facts \cite{holevo11,horn}
\begin{align}
||A B||_2 &\le ||A||\cdot ||B||_2,
\\
||V_{(q)}^{-1}|| &= 
||V_{(q)}^{-\top}|| = ||V_{(q)}^{-\top}V_{(q)}^{-1}||^{1/2} = 
||\Xi_{(q)}^{-1}||^{1/2},
\\
||\Theta_{(q),\mu}||_2^2 &= \sum_{r=0}^q \sum_{s=0}^q \bk{\delta^{r+s}_\mu}^2 
=\sum_{r=0}^q \sum_{s=0}^q \delta^{r+s}_\mu 
\le \mu+1.
\end{align}
Combining Eq.~(\ref{hs}), (\ref{zero_dL}), (\ref{alpha2}), and
(\ref{Q2}), I obtain
\begin{align}
||\tilde\Lambda_{,\mu}||_2^2 &\le \frac{\mu+1}{2}\Delta^{-2\mu}
 \sum_{q=\lceil\mu/2\rceil}^\infty \zeta_q,
\label{hsbound}
\\
\zeta_q &\equiv \frac{\phi_{2q}}{q!}\bk{\frac{\Delta}{w}}^{2q} ||\Xi_{(q)}^{-1}||^2.
\end{align}
Since $\Xi$ and therefore its submatrix $\Xi_{(q)}$ are
positive-definite \cite{horn}, $||\Xi_{(q)}^{-1}||$ is the largest
eigenvalue of $\Xi_{(q)}^{-1}$, which is equal to the inverse of the
smallest eigenvalue of $\Xi_{(q)}$.  Let $\lambda_q$ be the smallest
eigenvalue of $\Xi_{(q)}$. The right-hand side of Eq.~(\ref{hsbound})
converges and $\tilde\Lambda_{,\mu}$ is Hilbert-Schmidt if it passes
the ratio test
\begin{align}
\lim_{q\to\infty} \abs{\frac{\zeta_{q+1}}{\zeta_q}} = 
\lim_{q\to\infty} \frac{1}{q+1} \frac{\Delta^2}{w^2}
\frac{\phi_{2q+2}}{\phi_{2q}} \frac{\lambda_{q}^2}{\lambda_{q+1}^2}
< 1.
\label{test}
\end{align}
Two cases are of interest:
\begin{enumerate}[label=(\alph*)]
\item $f(x|\theta)$ is any probability density in the
  Szeg\H{o} class with compact support within $[x_1,x_2]$,
  $|x_j| < \infty$ \cite{widom66,berg11}, viz.,
\begin{align}
\mathcal S &\equiv \int_{x_1}^{x_2} dx \frac{\ln f(x|\theta)}
{\sqrt{(x-x_1)(x_2-x)}} > -\infty.
\label{szego}
\end{align}
For example, any strictly positive $f$ is in the class, as there
exists a $\delta$ such that $f \ge \delta > 0$ and
$\ln f \ge \ln \delta > -\infty$, leading to
\begin{align}
\mathcal S &\ge \ln\delta
\int_{x_1}^{x_2} \frac{dx}
{\sqrt{(x-x_1)(x_2-x)}} = \pi\ln\delta > -\infty.
\end{align}
If Eq.~(\ref{szego}) is satisfied, it is known \cite{widom66,berg11}
that, for $q \to \infty$, there exist constants $\Omega > 0$ and
$0 < \tau < 1$ such that
\begin{align}
\lambda_q &\to \Omega \sqrt{q} \tau^q,
&
\frac{\lambda_q^2}{\lambda_{q+1}^2} &\to \frac{1}{\tau^{2}}.
\label{asymptotic_lambda}
\end{align}
Furthermore, since $x^2 \le \max(|x_1|,|x_2|)^2$ for
$x \in [x_1,x_2]$,
\begin{align}
  \phi_{2q+2} &= \int_{x_1}^{x_2} dx f(x|\theta) x^{2q+2}
\nonumber\\
  &\le 
  \max(|x_1|,|x_2|)^2 \int_{x_1}^{x_2} dx f(x|\theta) x^{2q}
\nonumber\\
  &= \max(|x_1|,|x_2|)^2\phi_{2q}.
\end{align}
The left-hand side of Eq.~(\ref{test}) can therefore be bounded as
\begin{align}
\lim_{q\to\infty} \abs{\frac{\zeta_{q+1}}{\zeta_q}} &\le 
\lim_{q\to\infty}\frac{\Delta^2\max(|x_1|,|x_2|)^2}{(q+1)w^2\tau^2},
\label{test2}
\end{align}
which approaches zero and passes the ratio test given by
Eq.~(\ref{test}) for any $w$, $\Delta$, and $|x_j|$.  Beyond the
Szeg\H{o} class, Eq.~(\ref{test}) is also satisfied if
$\lambda_q^2/\lambda_{q+1}^2 = o(q)$, or if
$\lambda_q^2/\lambda_{q+1}^2 = O(q)$ and a small enough $\Delta/w$ is
chosen.

\item $f(x|\theta) \propto \exp(-x^2/2)$. Then the standard deviation
  of $F(X|\theta)$ is $\Delta$ and $\phi_{2q+2}/\phi_{2q} = 2q+1$.  It
  is known that \cite{szego36,berg11}
\begin{align}
\lambda_q &\to \Omega q^{1/4}\tau^{\sqrt{q}},
&
\frac{\lambda_q^2}{\lambda_{q+1}^2} &\to 1.
\label{asymptotic_lambda2}
\end{align}
Equation~(\ref{test}) becomes
\begin{align}
\lim_{q\to\infty} \abs{\frac{\zeta_{q+1}}{\zeta_q}} &= \frac{2\Delta^2}{w^2} < 1,
\end{align}
which is satisfied if 
\begin{align}
w > \sqrt{2}\Delta.
\end{align}
\end{enumerate}

\subsection{Summary}
To summarize, Appendix~\ref{sec_bounded} shows that $\tilde\Pi$ is
trace-class if $|\Psi(k)|^2$ is one of the following:
\begin{enumerate}[label=(\roman*)]
\item bandlimited with any choice of $w$, or
\item Gaussian with $w < 1/(\sqrt{2}\beta)$,
\end{enumerate}
while Appendix~\ref{sec_hs} shows that $\tilde\Lambda_{,\mu}$ is
Hilbert-Schmidt if $F(X|\theta)$ is one of the following:
\begin{enumerate}[label=(\alph*)]
\item in the Szeg\H{o} class with any choice of $w$, or
\item Gaussian with $w > \sqrt{2}\Delta$.
\end{enumerate}
Thus the choice of $w$ becomes an issue only if both are Gaussian.  To
satisfy both (ii) and (b), the standard deviations should satisfy
\begin{align}
\beta\Delta < \frac{1}{2},
\end{align}
such that a choice within $\sqrt{2}\Delta < w < 1/(\sqrt{2}\beta)$ is
possible.

Taking $\Delta \ll 1$, $\beta = O(1)$, and $w = O(1)$,
$||\tilde\Pi|| = O(1)$ and the right-hand side of Eq.~(\ref{hsbound})
converges to $O(\Delta^{-2\lfloor\mu/2\rfloor})$ under the conditions
above. Equation~(\ref{Kmumu2}) becomes
\begin{align}
\tilde K_{\mu\mu} &\le O(\Delta^{-2\lfloor\mu/2\rfloor}),
\end{align}
which is consistent with Eq.~(\ref{bound}).

With a trace-class $\tilde\Pi$, $\tilde\Lambda_{,\mu}$ is said to be
square-summable with respect to $\tilde\Pi$ if and only if
$\tilde K_{\mu\mu}$ converges \cite{holevo11}.  An operator is
guaranteed to be square-summable if it is bounded, and may still be so
even if it is unbounded \cite{holevo11}. As Hilbert-Schmidt operators
are a subclass of bounded operators, requiring $\tilde\Lambda_{,\mu}$
to be Hilbert-Schmidt may be an overkill; more relaxed conditions for
the convergence of $\tilde K_{\mu\mu}$ may exist. Choosing a different
scaling matrix $W$ can also lead to other conditions.

\section{\label{sec_convexity}Quantum bounds via convexity and
  classical simulation}
Discretize $F(X|\theta)$ as a distribution of point sources, such that
\begin{align}
F(X|\theta) &= \sum_{s} F_s \delta(X-X_s),
\\
\rho_1 &= \sum_{s} F_s e^{-i\hat kX_s}\ket{\psi}\bra{\psi} e^{i\hat kX_s}.
\end{align}
First assume that $\{F_s\}$ are known.  Denoting the QFI with respect
to parameters $\{X_s\}$ as $K^{(X)}$, I can use the convexity of QFI
\cite{fujiwara01,alipour,ng16} to write
\begin{align}
  K^{(X)}(\rho_1) &\le G,
  \\
G &\equiv \sum_s F_s K^{(X)}
\bk{e^{-i\hat kX_s}\ket{\psi}\bra{\psi} e^{i\hat kX_s}},
\\
G_{st} &= 4F_s \beta^2\delta_{s}^t,
\\
\beta &=
\sqrt{\bra{\psi}\hat k^2\ket{\psi}- (\bra{\psi}\hat k\ket{\psi})^2}.
\end{align}
With
\begin{align}
\theta_\mu &= \sum_s F_s X_s^\mu,
& H_{\mu s} &\equiv \parti{\theta_\mu}{X_s} = F_s \mu X_s^{\mu-1},
\end{align}
I can transform the Cram\'er-Rao bounds back to the ones with respect
to $\theta$ as
\begin{align}
K(\rho_1)^{-1} &\ge H G^{-1} H^\top, 
\\
\bk{H G^{-1} H^\top}_{\mu\nu}
&= \frac{\mu\nu \theta_{\mu+\nu-2}}{4\beta^2} = O(\Delta^{\mu+\nu-2}).
\end{align}
Hence
\begin{align}
[J(P)^{-1}]_{\mu\mu} &\ge \frac{[K(\rho_1)^{-1}]_{\mu\mu}}{N} \ge
\frac{(H G^{-1} H^\top)_{\mu\mu}}{N} 
\nonumber\\&
= \frac{\mu^2 \theta_{2\mu-2}}{4N\beta^2} 
=\frac{O(\Delta^{2\mu-2})}{N}.
\label{convexity}
\end{align}
The scaling of this bound with respect to $\Delta$ is looser than that
of the inverse of Eq.~(\ref{bound}) for $\Delta\ll 1$ and $\mu > 2$
but does not rely on the $\Delta\ll 1$ approximation.

Yet another bound can be computed by treating $\{F_s\}$ as parameters
and using the classical-simulation technique \cite{demkowicz}:
\begin{align}
K^{(F)}(\rho_1) &\le  J^{(F)}(F),
\\
J_{st}^{(F)}(F) &= \sum_u \frac{1}{F_u}\parti{F_u}{F_s}\parti{F_u}{F_t}
=\frac{\delta_{s}^t}{F_s},
\\
K(\rho_1)^{-1} &\ge R J^{(F)}(F)^{-1}R^\top,
\\
R_{\mu s} &\equiv \parti{\theta_\mu}{F_s} = X_s^\mu,
\\
\Bk{R J^{(F)}(F)^{-1}R^\top}_{\mu\nu} &=\theta_{\mu+\nu} = O(\Delta^{\mu+\nu}).
\end{align}
This proof is a straightforward generalization of Appendix~D in
Ref.~\cite{tsang_semiclassical}.  The final result is
\begin{align}
[J(P)^{-1}]_{\mu\mu} &\ge \frac{[K(\rho_1)^{-1}]_{\mu\mu}}{N} \ge
\frac{[R J^{(F)}(F)^{-1}R^\top]_{\mu\mu}}{N} 
\nonumber\\&
= \frac{\theta_{2\mu}}{N} = \frac{O(\Delta^{2\mu})}{N},
\end{align}
the scaling of which is unfortunately looser than those of
Eqs.~(\ref{bound}) and (\ref{convexity}) for $\Delta \ll 1$.

\section{\label{sec_prefactor}Decay of the quantum SNR
for higher moments}
With Eq.~(\ref{Keven}), the quantum SNR given by Eq.~(\ref{snr}) for
an even $\mu = 2q$ can be expressed in terms of the normalized
quantities defined by Eqs.~(\ref{theta_phi})--(\ref{Lambda_V}) as
\begin{align}
\textrm{QSNR}_{2q}
  &=  N\Bk{\chi_{q}\Delta^{2q} + o(\Delta^{2q})},
\\
\chi_{q} &= \frac{\bra{\psi}\hat k^{2q}\ket{\psi}\phi_{2q}^2}
{q!^2 (V_{qq})^2},
\label{prefactor}
\end{align}
where $\phi_{2q}$ is a normalized object moment and $V$ is the
Cholesky factor of the normalized Hankel matrix $\Xi$. For a given
$\Delta$ in the subdiffraction regime, the SNR as a function of $q$
depends on not only $\Delta^{2q}$ but also the prefactor $\chi_q$. Here I
show that the sequence $\{\chi_q: q \in \mathbb N\}$ is bounded under
benign conditions, so the SNR must decay with $q$ at least as quickly
as the exponential $\Delta^{2q}$.

If the OTF is bandlimited or Gaussian with bandwidth $\beta < \infty$,
Eqs.~(\ref{otfbound1}) and (\ref{otfbound2}) give
\begin{align}
\bra{\psi}\hat k^{2q}\ket{\psi} &\le \frac{(2q)!}{q!2^q}\beta^{2q}.
\label{otfbound3}
\end{align}
If the $f(x|\theta)$ given by Eq.~(\ref{f}) has a compact support
within $[-1,1]$,
\begin{align}
\phi_{2q} &\le 1.
\label{phibound}
\end{align}
As the support has been assumed to contain an infinite number of
points, $\Xi_{(q)} > 0$, and $V_{qq} > 0$ is an
eigenvalue of the lower-triangular Cholesky factor $V_{(q)}$
\cite{horn}. Let $v$ be the eigenvector of $V_{(q)}$
with eigenvalue $V_{qq}$ and $v^\top v = 1$. 
Then 
\begin{align}
V_{qq}^2 = v^\top V_{(q)}^\top V_{(q)}v \ge \min_{v^\top v=1}
v^\top V_{(q)}^\top V_{(q)}v = \lambda_q,
\label{Vbound}
\end{align}
where $\lambda_q$ is the smallest eigenvalue of $V_{(q)}^\top V_{(q)}$
and also $V_{(q)}V_{(q)}^\top=\Xi_{(q)}$ \cite{horn}, so
$\lambda_q > 0$.  Substituting Eqs.~(\ref{otfbound3})--(\ref{Vbound})
into Eq.~(\ref{prefactor}) gives
\begin{align}
\chi_q&\le \frac{(2q)!\beta^{2q}}{q!^3 2^q\lambda_q}
\equiv \chi_q'.
\end{align}
$\chi_q' < \infty$ for any finite $q$, and if $f(x|\theta)$ is in the
Szeg\H{o} class, $\lambda_q$ obeys Eqs.~(\ref{asymptotic_lambda}) as
$q \to \infty$, leading to $\lim_{q\to\infty} \chi_q' = 0$.  Hence
$\{\chi_q': q \in \mathbb N\}$ is a bounded sequence, so is
$\{\chi_q: q \in \mathbb N\}$, and there exists a finite $\tilde\chi$
such that
\begin{align}
  \chi_{q} &\le \tilde\chi < \infty,
  &
\textrm{QSNR}_{2q} &\approx N \chi_q \Delta^{2q} \le N \tilde\chi \Delta^{2q}.
\end{align}
A similar decay behavior of the quantum SNR for the odd moments can be
shown via the same procedure.

\section{\label{sec_zhou}Review of Ref.~\cite{zhou18v2}}
Here I summarize the essential arguments in Ref.~\cite{zhou18v2},
using the notations and parametrization here and focusing on the
one-photon state for simplicity. Rewrite Eq.~(\ref{expand}) as
\begin{align}
\rho_1 &= \sum_{\nu=0}^\infty \theta_\nu \sigma_\nu,
\\
\sigma_\nu &\equiv \sum_{q=0}^\nu 
\frac{(-i\hat k)^q\ket{\psi}\bra{\psi}(i\hat k)^{\nu-q}}
{q!(\nu-q)!},
\end{align}
such that the probability distribution for a measurement $E_1(\xi)$
obeys
\begin{align}
\pi(\xi|\theta) &= \trace E_1(\xi)\rho_1 = \sum_{\nu=0}^\infty \theta_\nu S_\nu(\xi),
\label{pi}
\\
S_\nu(\xi) &\equiv \trace E_1(\xi)\sigma_\nu = 
\sum_{q=0}^\nu \frac{\bra{\psi}(i\hat k)^{\nu-q}E_1(\xi)(-i\hat k)^q\ket{\psi}}
{q!(\nu-q)!}.
\end{align}
The Fisher information for $\theta_\mu$ becomes
\begin{align}
J_{\mu\mu} &= N \sum_\xi \frac{\pi_{,\mu}(\xi|\theta)^2}{\pi(\xi|\theta)}
=N\sum_\xi \frac{S_\mu^2(\xi)}{\pi(\xi|\theta)}
\nonumber\\&
= N \Delta^{-\mu} \sum_\xi \frac{\Delta^\mu |S_\mu(\xi)|}{\pi(\xi|\theta)}|S_\mu(\xi)|.
\end{align}
If
\begin{align}
\frac{\Delta^\mu |S_\mu(\xi)|}{\pi(\xi|\theta)} &\le c_\mu = O(1),
&
\sum_\xi |S_\mu(\xi)| &\equiv d_\mu < \infty,
\end{align}
then
\begin{align}
J_{\mu\mu} &\le c_\mu d_\mu N\Delta^{-\mu} = NO(\Delta^{-\mu}),
\label{zhou1}
\end{align}
which is essentially Theorem 1 in Ref.~\cite{zhou18v2}.  To prove
$c_\mu = O(1)$, note that $S_\mu(\xi) \neq 0$ must hold for
$\pi_{,\mu}(\xi|\theta) \neq 0$, so the expansion in Eq.~(\ref{pi})
must contain at least the term $\theta_\mu S_\mu(\xi)$. In other
words,
\begin{align}
\pi(\xi|\theta) &= O(\Delta^\alpha), & \alpha &\le \mu.
\end{align}
Coupled with the proof of $|S_\mu(\xi)| < \infty$ in
Ref.~\cite{zhou18v2} and the fact $\pi(\xi|\theta) > 0$,
\begin{align}
  \frac{\Delta^\mu |S_\mu(\xi)|}{\pi(\xi|\theta)} = 
  \frac{\Delta^\mu |S_\mu(\xi)|}{O(\Delta^\alpha)} = 
  O(\Delta^{\mu-\alpha}).
\end{align}
Reference~\cite{zhou18v2} also proves $d_\mu < \infty$ under
reasonable conditions.

Compared with Eqs.~(\ref{Keven}), (\ref{Kodd}), and (\ref{bound}), not
only is the scaling of Eq.~(\ref{zhou1}) with $\Delta$ for odd moments
less tight, the value of its prefactor $c_\mu d_\mu$ also depends on
the measurement and does not seem easy to compute. Without a more
concrete prefactor, it would not be possible to study the SNR as a
function of $\mu$ for a given $\Delta$ like
Appendix~\ref{sec_prefactor} and show that higher moments are more
difficult to estimate, as the prefactor may increase quickly with
$\mu$.


Reference~\cite{zhou18v2} further argues that the optimal POVM that
maximizes the Fisher information for a given $\theta_\mu$ should
satisfy
\begin{align}
E_1(\xi)(-i\hat k)^q\ket{\psi} &= 0
\textrm{ for }q < \floor{\frac{\mu}{2}},
\label{nulling}
\end{align}
in order to obtain
\begin{align}
S_\nu(\xi) &= 0 \textrm{ for }\nu < 2\floor{\frac{\mu}{2}},
&
\pi(\xi|\theta) &= O(\Delta^{2\lfloor\mu/2\rfloor}).
\end{align}
This leads to
\begin{align}
 \max_{E_1}J_{\mu\mu}(\theta) \stackrel{?}{=} NO(\Delta^{-2\lfloor\mu/2\rfloor}),
\label{zhou3}
\end{align}
which is essentially their Theorem 3. This argument seems to be flawed
however: it is not clear that Eq.~(\ref{nulling}) is a necessary
condition for the optimal POVM.  Although it leads to a scaling that
is close to the one suggested by Eq.~(\ref{zhou1}), the scaling is not
the only concern when evaluating $\max_{E_1}J_{\mu\mu}(\theta)$ at a
specific $\theta$; the prefactor also matters. There may exist a POVM
that violates Eq.~(\ref{nulling}) and obeys a worse overall scaling
but gives a prefactor large enough to make the information higher at
that specific $\theta$. This would imply that the optimal POVM does
not satisfy Eq.~(\ref{nulling}), and Eq.~(\ref{zhou3}) does not follow
from Eq.~(\ref{nulling}).


\bibliography{research}

\end{document}